\documentclass[aps,prb,reprint,groupedaddress]{revtex4-1}

\usepackage{amsmath}
\usepackage{graphicx}
\usepackage{bm}
\usepackage{comment}
\usepackage{makecell}
\usepackage{url}
\usepackage{hyperref}
\usepackage{natbib}

\usepackage{wasysym}

\begin{document}


\title{Spin pumping efficiency in room-temperature CdSe nanocrystal quantum dots}

\author{A. Khastehdel Fumani, J. Berezovsky}
\email[Corresponding author:]{jab298@case.edu}
\affiliation{Department of Physics, Case Western Reserve University, Cleveland, Ohio 44106}
\date{\today}
\begin{abstract}
 To understand and optimize optical spin initialization in room temperature CdSe nanocrystal quantum dots (NCQDs) we studied the dependence of the time-resolved Faraday rotation signal on pump energy $E_p$ in a series of NCQD samples with different sizes. In larger NCQDs, we observe two peaks in the spin signal vs. $E_p$, whereas in smaller NQCDs, only a single peak is observed before the signal falls to a low, broad plateau at higher energies.  We calculate the spin-dependent oscillator strengths of optical transitions using a simple effective mass model to understand these results. The observed $E_p$ dependence of the spin pumping efficiency (SPE) arises from the competition between the heavy hole (hh), light hole (lh) and split-off (so) band contributions to transitions to the conduction band.  The two latter contributions lead to an electron spin polarization in the opposite direction from the former.  At lower $E_p$ the transitions are dominated by the hh band, giving rise to the low energy peaks. At higher $E_p$, the increasing contributions from the lh and so bands lead to a reduction in SPE.  The different number of peaks in larger and smaller NCQDs is attributed to size-dependence of the ordering of the valence band states. \end{abstract}
\maketitle

 Electrons confined in quantum dots (QDs) are a useful system for studying the physics of single spins which may lead to applications in spintronics or quantum information processing.\cite{Spintronics_Awschalom,Berezovsky_Book} Unlike gate-defined\cite{SingleElectronGate} or self-assembled\cite{Gammon} QDs which confine electrons on 10-100 nm length-scales, colloidal nanocrystal QDs provide much stronger confinement, on length-scales of a few nanometers. This strong confinement has the desirable property of enabling operation at room temperature and above, but also causes the spin physics to deviate significantly from bulk-like properties.   

 Optical excitation of spins in semiconductors is a powerful technique to study the initialization of coherent spin states in both bulk materials and nanostructures. Here, we study the room temperature optical spin pumping process in nanocrystal QDs (NCQDs).  We measure the efficiency of spin pumping vs. excitation energy in NCQD ensembles with different mean particle size and find that the resulting spectra depend on the confinement-induced valence (v) band mixing and the size-dependent ordering of v-band states.  In larger, low-temperature QD systems, these effects are present to an extent,\cite{Berezovsky_Book} but the smaller size of the NCQDs studied here brings confinement-induced effects into a dominant role.  These results reveal the spin-dependent optical selection rules in NCQDs, and can be applied to optimize optical spin initialization at room temperature in these materials.

\begin{figure}[h]
	\centering
	\includegraphics[scale=0.65] {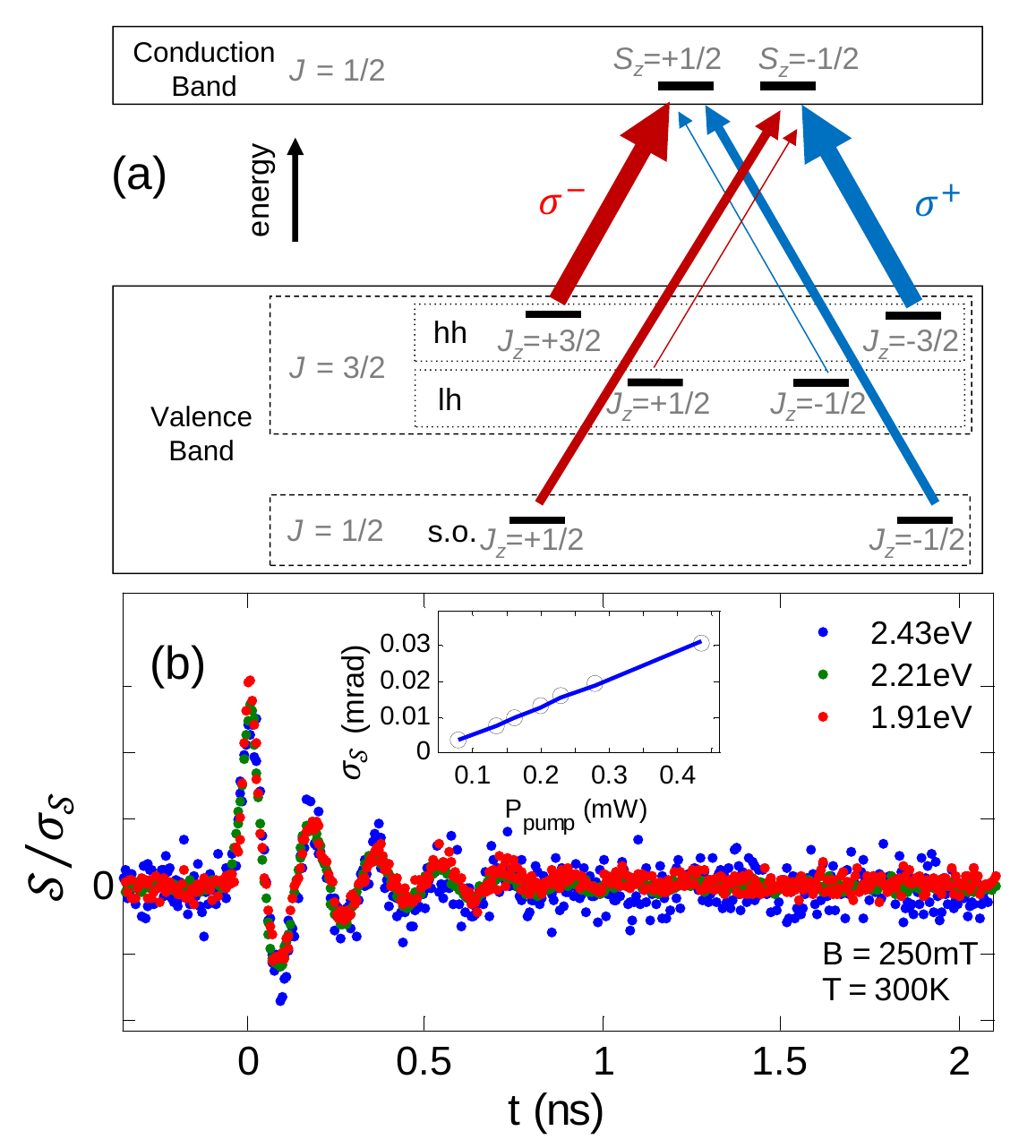}
	\caption{(a)Schematic of possible optical transitions through $\sigma^+$ or $\sigma^-$ light from hh, lh, and s.o. to c-band. (b)$\mathcal{S}(t)$ from sample A normalized by $\sigma_{\mathcal{S}}$ for three different $E_p$. The inset shows $\sigma_{\mathcal{S}}$ vs $P_{pump}$ at $E_p=1.97$~eV along with a linear fit.}
	\label{fig_EXP}
	\end{figure}

 Optical spin pumping in bulk semiconductors can be achieved by exciting electrons using circularly polarized light into the conduction (c) band from a particular valence sub-band (heavy hole (hh), light hole (lh), or split-off hole (s.o.)).~\cite{meier1984}  A photon carrying an angular momentum e.g. $-1\hbar$ can excite an electron from the hh sub-band to the c-band, creating an electron-hole pair with $ | J_{z,h}=+3/2,S_{z,e}=+1/2 \rangle$ or from the lh or s.o. band creating an electron-hole pair with $|J_{z,h}=+1/2,S_{z,e}=-1/2 \rangle$ (see Fig.\ref{fig_EXP}~(a)).  Here $J_{z,h}$ and $S_{z,e}$ are the v-band and c-band angular momentum projections, respectively. In bulk semiconductors, the hole spin lifetime ($\tau_h <1$~ps)\cite{Dyakonov_text} is typically much shorter than that of the electron.  On timescales longer than $\tau_h$, the electron spin in the c-band is the total net spin in the system.  In this situation, transitions from the lh and s.o. bands produce spin polarization opposite to that produced from the transitions from the hh band for the same circularly polarized excitation. The net spin pumped into the c-band will therefore depend on the excitation energy and the relative oscillator strength of transitions from the three valence sub-bands. The oscillator strengths of transitions from the hh, lh, and s.o. bands to the c-band are in the ratio of 3:1:2.~\cite{meier1984}   Thus at an excitation energy where the hh and lh bands are excited equally, net spin polarization is still created in the c-band. If all three sub-bands are excited equally, however, zero net c-band spin polarization is generated.

 The process of optical spin pumping in QDs differs from bulk semiconductors.  In QDs, the hole eigenstates are composed of a superposition of bulk eigenstates from all three valence sub-bands in general.  This v-band mixing depends on the quantum confinement.  In larger quantum dots, the hole states are still often dominated by one of the valence sub-bands, and thus resemble the bulk case.  In the strong confinement in the NCQDs studied here, the hole eigenstates have substantial sub-band mixing, and even the ordering of the eigenstates changes with QD size.~\cite{Ekimov93}

 Spin dynamics and relaxation processes in CdSe NCQDs have been subject of previous studies\cite{Stern_SpinElectrochemically_PRB2005,Gupta_SpinDynamics_PRB2002,Fumani,Scholes_ExcitonSpinRelaxation_2006}, however the details of the optical spin pumping process studied here have not been the focus of any prior work. The results presented here reveal how excitation from different quantum-confined v-band levels in various sizes of NCQDs affects the resultant spin polarization.  We understand the experimental results by comparing them to a simple, spherical effective mass model (following Ref.~\onlinecite{Ekimov93} and \onlinecite{Grigoryan}).  We find that in larger NCQDs, there are several low energy transitions dominated by the hh band, resulting in efficient spin pumping.  In smaller NCQDs, we find less efficient spin pumping, as the stronger quantum confinement pushes the energy of many of the hh-dominated transtions above that of lh and s.o.-dominated transitions.

 We performed a series of time resolved Faraday rotation (TRFR) measurements\cite{Crooker_PRL96} on colloidal solutions of CdSe NCQDs at room temperature. The NCQDs which are stabilized with octadecyl amine surfactants are suspended in toluene with a concentration of approximately 2~mg/mL (purchased from NN-Labs). We studied three samples with different nanocrystal sizes with the first absorption peaks of the ensembles centered at energies $E_{abs}=1.94$, 2.07, and 2.10~eV. They are referred to as sample A, B, and C respectively hereafter. From the calculations discussed below, we find that these $E_{abs}$ correspond to mean NCQD radii  $a_A=3.05$, $a_B=2.40$, and $a_C=2.04$~nm. Size distribution measurement of sample A using transmission electron microscopy (TEM) confirms this size assignment.\cite{Fumani} TEM measurements also reveal that the NCQD ensemble has a normal size distribution with $10$\% fractional width. We assume the same width of size distribution for the other samples.

 In the pump/probe TRFR measurement carried out here, a circularly polarized pump and a linearly polarized probe laser are focused to the same spot inside a 1~mm thick cuvette containing the sample. The cuvette is placed between poles of an electromagnet which provides transverse magnetic field $B$. The pump pulse creates a net spin polarization in the sample. This spin polarization modifies the index of refraction for right and left handed circularly polarized light. Consequently a linearly polarized probe pulse sent through the same spot in the sample measures a projection of the induced spin polarization through a rotation in its polarization axis. We measure this rotation through a two-lock-in modulation technique. Both pump and probe pulses are derived from a pulsed supercontinuum laser at 5~MHz repetition rate, and have duration of $\approx 20$~ps. We used two sets of linearly graded edgepass filters to independently select the wavelengths from the white light output of the laser as pump and probe beams. The probe pulse is fixed at an energy slightly less than $E_{abs}$, and has bandwidth $\Delta \lambda = 30$~nm.  The pump pulse has bandwidth $\Delta\lambda = 20$~nm ($\Delta E = 0.06\mbox{-}0.1$~eV across the energy range studied here).    A mechanical delay line was used to adjust the pump-probe delay, mapping out time evolution of the spin polarization projected onto the probe pulse propagation direction.

 Using the wide range energy output of the supercontinuum laser we were able to scan the pump photon energy $E_p$ from $1.8$~eV to $2.7$~eV and study the resulting spin polarization. Figure~\ref{fig_EXP}(b) shows room temperature TRFR data from sample A taken at three different $E_p$ at $B=250$~mT. The arrival of pump pulse at $t=0$ creates an electron-hole pair, which rapidly relaxes to the lowest energy exciton state.\cite{EnergyRelaxation_Woggon}  The hole spin state is randomized on a timescale faster than we can observe here,\cite{Scholes_ExcitonSpinRelaxation_PCB2005, Scholes_ExcitonSpinRelaxation_2009} and the resulting electron spin polarization is seen as the first spike in the data. Precession of spins around the external magnetic field $B$ in the Voigt geometry is mapped as oscillations in the resultant signal.  The magnetic-field-dependent oscillations, which change sign with opposite excitation helicity, prove that the signal arises from spins in the QDs.  The spin signal decays in the course of a few nanoseconds because of decoherence and dephasing processes.\cite{Zutic_Spintronic_RMP2004} 

 To quantify and analyze spin pumping efficiency (SPE) at different $E_p$ we consider the measured FR signal at time $t$ denoted as $\mathcal{S}(t)$, a summation of spin signal and a noise term: $ \mathcal{S}(t)=  s(t)+  n(t)$.  Assuming Gaussian uncorrelated noise with standard deviation $\sigma_n$, we write the standard deviation of the signal as $\sigma_{\mathcal{S}} = \sqrt{ \sigma_s^2 + \sigma_n^2}$. Given that we can measure the noise level $\sigma_n$ at times prior to arrival of the pump pulse, we calculate and use $\sigma_{\mathcal{S}}$ as a measure of SPE.

 To demonstrate that $\sigma_{\mathcal{S}}$ is a proper measure of SPE three different FR curves taken at distinct $E_p$ are normalized by $\sigma_{\mathcal{S}}$ and shown in Fig.~\ref{fig_EXP}(b). The raw data from which the superposed curves are derived have different amplitudes originally, as they are taken at different $E_p$ and different pump power $P_{pump}$.  The fact that this normalization collapses the curves onto each other signifies that $\sigma_{\mathcal{S}}$ is a good measure of the spin signal amplitude.  Furthermore, the fact that the observed spin dynamics are the same across the range of $E_p$ indicates that changing $E_p$ only changes the degree of spin initialization, not the resulting dynamics.  The inset to Fig.~\ref{fig_EXP}(b) shows the dependence of $\sigma_s$ on $P_{pump}$, which is well-described by a linear fit.  As $P_{pump}$ is not constant across the laser spectrum, the SPE data discussed below will be normalized by $P_{pump}$, which is kept within the linear regime shown in the inset. 

\begin{figure}[htbp]
	\centering
	\includegraphics[scale=0.55] {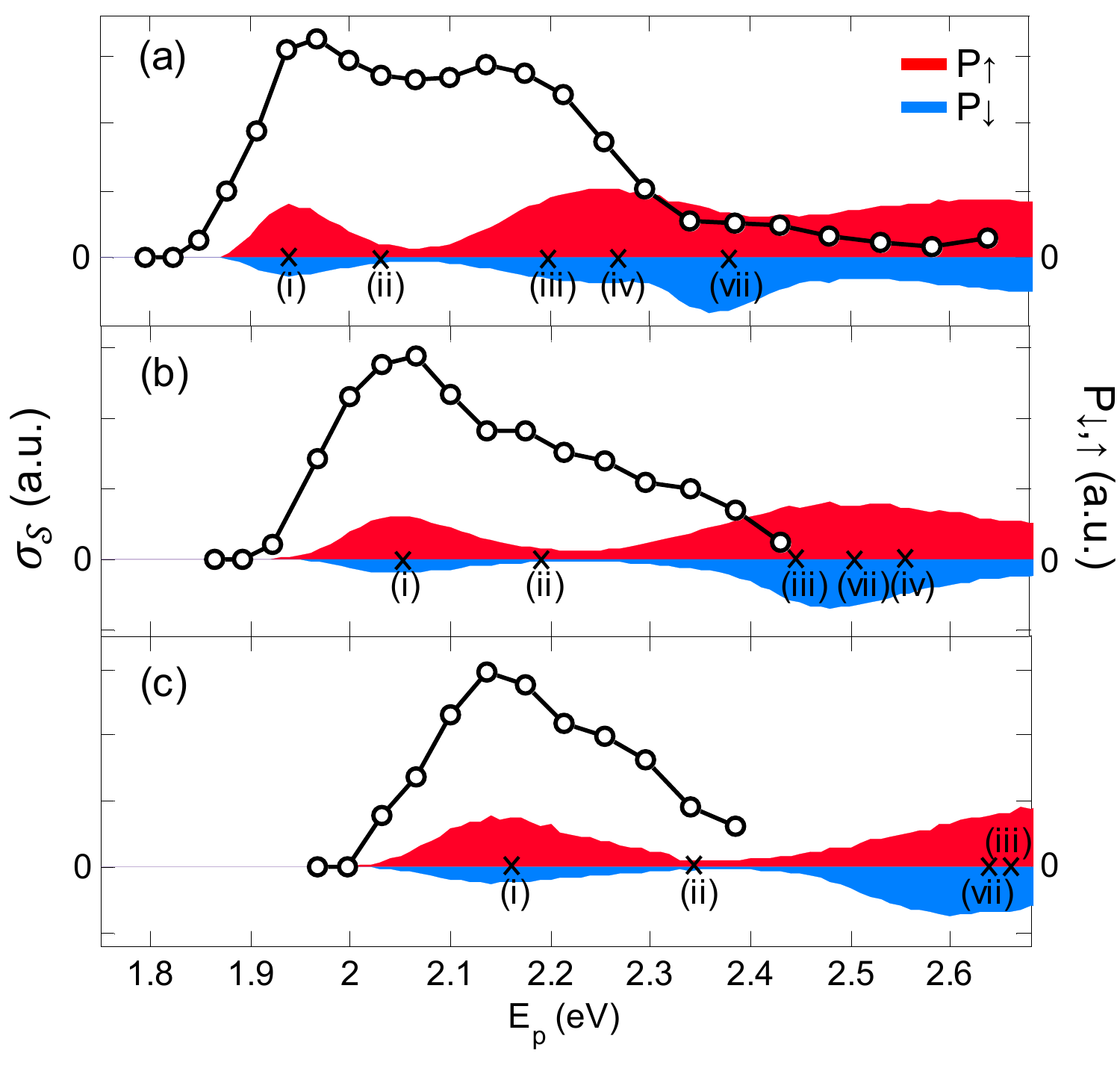}
	\caption{(a)-(c) show $\sigma_{\mathcal{S}}$ vs. $E_p$ for sample (A)-(C) respectively. The Red(blue) regions represent $P_{\uparrow}$($P_{\downarrow}$) vs. $E_p$. Labels on $E_p$-axes show important transitions as named in Fig.~\ref{fig_energy}}
	\label{fig_efficiency}
	\end{figure}

 Figures \ref{fig_efficiency}(a)-(c) show $\sigma_s$ vs. $E_p$, yielding the SPE spectrum in all three samples (black circles).  Each sample shows a low energy peak that corresponds roughly to $E_{abs}$.  Sample A, with the largest NCQD size, shows a second peak approximately 200~meV higher. At higher energies SPE falls off to a low broad plateau. In contrast, in samples B and C no clear second peak is observed in the SPE spectrum, but instead a broad shoulder is seen as $E_p$ is increased.  These results can be understood by looking at the spin dependent optical transition strengths of NCQDs and size-dependence of electron and hole energy levels. 

 The key to interpreting the data is how an admixture of hh, lh, and so valence sub-bands are involved in excitations with different energies. Using an effective mass model, following the work of Ekimov et al,\cite{Ekimov93} we calculated the eigenstates and energies of conduction and v-band states and the relevant optical transition strengths in spherical CdSe nanocrystals.

 We first calculated the quantum-confined electron and hole wavefunctions with appropriate boundary conditions. The holes are considered to be bound inside the NCQD within an infinite potential well while the electron wavefunction can leak out.~\cite{NorrisBawendi_PRB1996} In general, the electron wavefunction can be written as: $\Psi^{e}_{l,m}(\mathbf{r})= f^c_{l,m}(\mathbf{r})u_{c,S_z}$. Where $u_{c,S_z}$ and $f^c_{l,m}$ are the Bloch and envelope functions in c-band. The subscripts $l$ and $m$ denote the envelope function angular momentum magnitude and projection quantum numbers.  By applying the boundary conditions, the radial part of the envelope function is determined to be spherical and modified Bessel functions for the inside and outside regions. Similarly the hole wavefunction is
	\begin{equation}
	\label{eq:hWF}
	\Psi^{h,\pm}_{F,M}(\mathbf{r})=
	\sum\limits_{l}
	\sum\limits_{J,J_z}
	f^{v,\pm}_{l,m}(\mathbf{r}) u_{J,J_z}
	\end{equation}
with $u_{J,J_z}$ and $f^{v,\pm}_{l,m} $ being the Bloch and envelope functions in v-band. 
The set of good quantum numbers for hole eigenstates are $F$ and $M$, where $F$ is the sum of angular momentum of the Bloch and envelope functions and $M = J_z +m$ is its projection. The $\pm$ superscript denotes even and odd parity states. For even states $l=F+1/2$ and $F-3/2$ and for odd states, $l=F-1/2$ and $F+3/2$. In the v-band, the radial part of the envelope function is in general a sum of three spherical Bessel functions $j_l(k_t r)$, with $k_t$ being the wavevectors in lh, hh, and s.o. sub-bands. The second summation in Eq.~\ref{eq:hWF} includes contributions from the three valence sub-bands. With $J=3/2$, the sum is over $J_z=\pm3/2,\pm1/2$ which includes the Bloch functions, $u_{3/2,J_z}$ of the hh and lh sub-bands. With $J=1/2$, the summation is over $J_z=\pm1/2$ which includes $u_{1/2,\pm1/2}$, the Bloch functions for the s.o. sub-band.  The energies and envelope functions are determined by imposing the appropriate boundary conditions.  The exciton energy levels are determined by the difference between c-band and v-band energies, with the electron-hole Coulomb energy included empirically, following Ref.~\onlinecite{NorrisBawendi_PRB1996}. The strength of a transition induced by right handed or left handed circularly polarized light $\sigma^{ \pm}$ is $P^{ \pm}= | \langle \Psi^{h}  | \mathbf{ e^\pm \hat{p}} |  \Psi^{e}\rangle|^2$. $\mathbf{e^\pm =e_x}\pm  i\mathbf{e_y}$ is the polarization vector and  $\mathbf{\hat{p}}$ is the momentum operator.  We calculate the transition strength in the envelope function approximation as
	\begin{equation}
	\label{eq:WholeSpin}
	P^{\pm}_{F,M}=
	\sum_{l,J,J_z}
	\int_{-\infty}^{\infty}
	dr f_{l^{'},m{'}}^c(\mathbf{r})f_{l,m}^v(\mathbf{r})
	|\langle u_{J,J_z} |
	\mathbf{e^\pm \hat{p}}|
	u_{c,S_z}
	\rangle|^2.
	\end{equation}

\begin{figure}[htbp]
	\centering
	\includegraphics[scale=0.6] {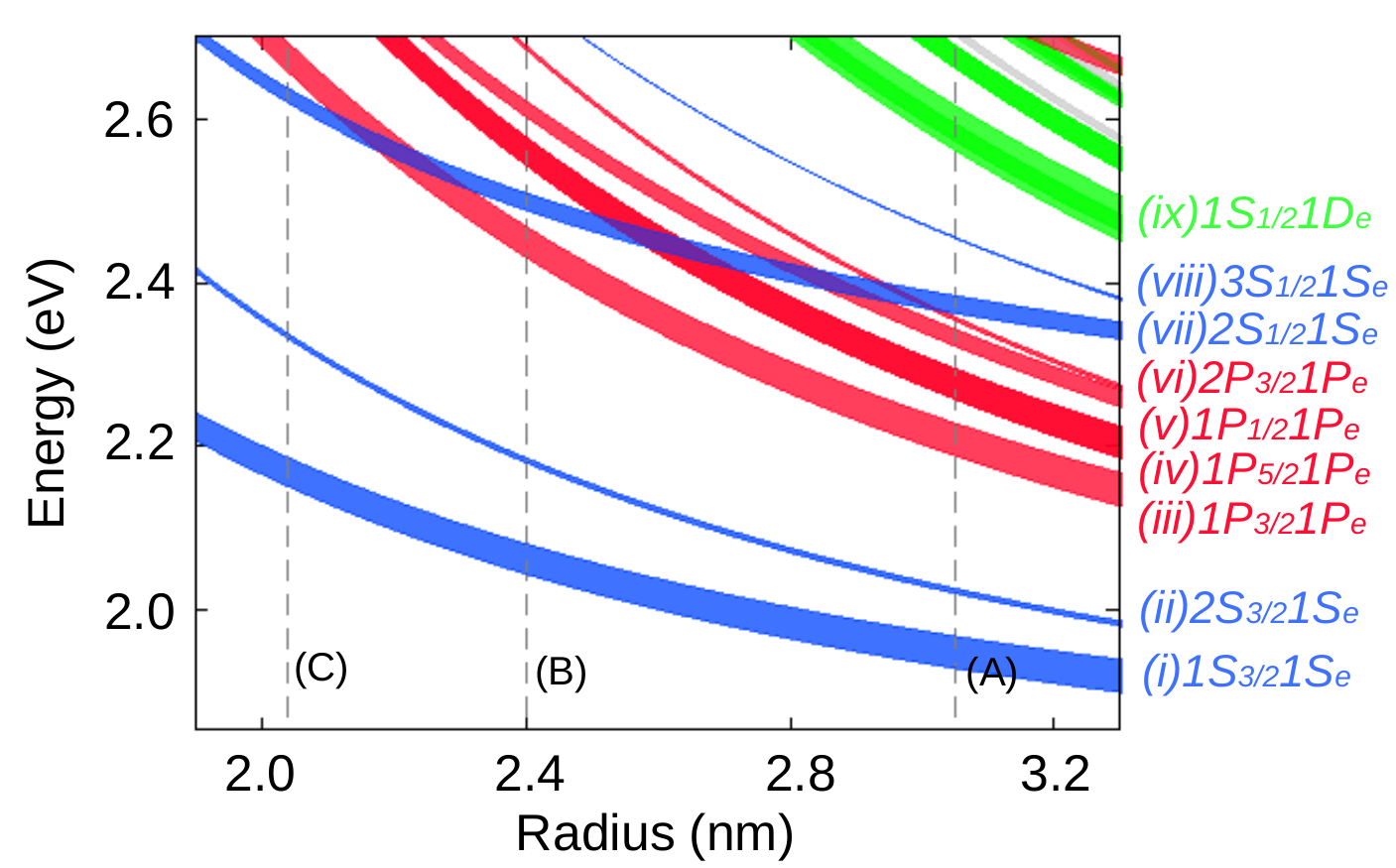}
	\caption{Calculated exciton energies vs. NCQD radius. Labels on right indicate the corresponding hole and electron states. The vertical dashed lines (A), (B), and (C) show the center radius of the size distribution of the respective samples.}
	\label{fig_energy}
	\end{figure}

 Figure~\ref{fig_energy} shows the calculated energy of optical transitions vs. radius $a$. The thickness of each line represents the strength of the particular transition.  (Only transitions with strength greater than 5\% of transition (i) are shown.) The transitions are labeled using the standard notation.\cite{Lipari_PRB1973}  First the v-band state is denoted $nl_F$, where $n$ is the principal quantum number, $l$ is the lowest orbital momentum included in the state.  The c-band state is then denoted $nl_e$. 

 The color of the lines in Fig.~\ref{fig_energy} represents $l$ of the c-band state (blue: S, red: P, and green: D).  For all sizes, the lowest transition is $1S_{3/2}1S_e$, and the next lowest is the relatively weak $2S_{3/2}1S_e$.   All of the transitions show decreasing energy with increasing radius, as the quantum confinement is reduced.   Notably, however, the transition energies do not all change at the same rate with radius.  For example, the $2S_{1/2}1S_e$ transition crosses other transitions as the NCQD radius changes.  For smaller NCQDs, (e.g. sample C, indicated by vertical dashed line (C) in Fig.\ref{fig_energy}), the third-lowest energy transition is $2S_{1/2}1S_{e}$.  At higher energies, four transitions involving holes in $1P_{3/2}$, $1P_{5/2}$,$1P_{1/2}$ , and $2P_{3/2}$ states and $1P_e$ electrons emerge. In contrast for larger NCQDs this cluster of $P$ transitions occurs at lower energies than level $2S_{1/2}1S_{e}$ (e.g. sample A). This change in the order of transitions has significant implications for optical spin pumping of large and small NCQDs.
	
 The fact that the hole wave function is a superposition of hh, lh, and so sub-bands, represented as the summation over $u_{J,J_z}$, means that e.g. if the transition $1S_{3/2}1S_e$ is driven it is not a transition purely from the hh sub-band to the c-band. In consequence, both c-band spin up and spin down are generated by each circularly polarized light polarity. The optical transitions shown in Fig.~\ref{fig_energy} include contributions from all three valence sub-bands, regardless of the c-band spin state they involve. However, to determine the probability of creating a certain c-band spin polarization by a circularly polarized photon we separately consider the oscillator strength of transitions to states with c-band spin up and spin down: $P^{\pm}_{\uparrow} = P^{\pm}(S_z=1/2)$ and spin down $P^{\pm}_{\downarrow} = P^{\pm}(S_z=-1/2)$.
 
 The result is depicted in Fig.~\ref{fig_efficiency} as the blue and red regions, which show histograms of $P^{+}_\uparrow(E_p)$ and $P^{+}_\downarrow(E_p)$ respectively for an ensemble of NCQDs with normally distributed radii with a fractional standard deviation of $10\%$.  The distribution of radii displays the effect of inhomogeneous broadening -- features from transitions with a larger slope of energy vs. radius will appear broader and with smaller amplitude than those with smaller slope. 

 Despite the simplicity of the model used, the patterns predicted by the calculations outlined above and depicted in histograms of Fig.~\ref{fig_efficiency} give qualitative physical insight to the experimental data. When the NCQDs are pumped resonantly, at $1.95$, $2.05$, and $2.15$~eV in samples A, B, and C respectively, the first peak in SPE occurs. The resonant transition is dominated by the hh sub-band and leads to up c-band spins.  In sample A, a cluster of transitions occurs (iii-vi), which are also dominated by the hh component.  These give rise to the second peak in Fig.~\ref{fig_efficiency}(a).  At higher energy, transition (vii) is excited, which is the first transition to be dominated by lh and s.o. hole components.  This transition excites mainly spin down in the c-band, leading to a sharp drop in the SPE.  At higher energies, transitions become closely spaced and contain all three valence sub-bands, leading to a low broad feature in the SPE spectrum.  In sample C, the smallest size NCQDs, the ordering of transition (vii) and the cluster (iii-vi) is reversed.  Thus the spectrum shows only the resonant peak and a small shoulder from the weaker transition (ii) before being cut off by the contributions form the lh and s.o. sub-band in transition (vii). The intermediate sample B, shows a broader shoulder, with some spin signal emerging from transition (iii) before the emerging significance of the lh and s.o. bands reduces the SPE. 

 The calculations predict additional broad, low peaks in the SPE at even higher energies.  For example in sample A, the calculation shows a low peak around 2.55~eV, where we instead observe just a broad shoulder.  This may indicate that the spin state is not perfectly preserved in the energy relaxation process for electron hole pairs excited to higher energies. Moreover the simple model used here does not take into account the lack of spherical symmetry which would allow additional transitions and can potentially smear out some of the calculated features.

 In summary, we have measured room-temperature SPE vs. excitation energy in three ensembles of NCQDs with varying sizes. In all samples, we find a peak in SPE at the energy resonant with the lowest-energy interband transition.  In larger size NCQDs, we observe a second peak in SPE about 200~meV above the first peak.  In contrast, we do not see the second peak in the two smaller samples.  These observations are understood by comparison to effective mass calculations, which show that these features can be explained by the size-dependent ordering of v-band states and mixing of valence sub-bands.  These results provide a picture of how strong quantum confinement affects spin-dependent optical transitions, and also allow for the optimization of optical spin pumping for future investigation of spin physics in these nanostructures.

\begin{acknowledgments}
This work was supported by AFOSR, award No. FA9550-12-1-0277.
\end{acknowledgments}

\bibliography{RefsSpinPumping}

\end{document}